\documentclass{article}

\usepackage[preprint]{neurips_2025}

\usepackage[utf8]{inputenc} 
\usepackage[T1]{fontenc}    
\usepackage{hyperref}       
\usepackage{url}            
\usepackage{booktabs}       
\usepackage{amsfonts}       
\usepackage{nicefrac}       
\usepackage{microtype}      
\usepackage{xcolor}         

\usepackage{graphicx}
\usepackage{float} 
\usepackage{subfigure}
\usepackage{algorithm}
\usepackage{algorithmic} 
\usepackage{amsmath}

\title{PIPO: Pipelined Offloading for Efficient Inference on Consumer Devices}

\author{
  Yangyijian Liu\textsuperscript{1} \quad Jun Li\textsuperscript{1} \quad Wu-Jun Li\textsuperscript{1} \\
  \textsuperscript{1} School of Computer Science, Nanjing University, China \\
  \texttt{\{yyj.liu, mg1933036\}@smail.nju.edu.cn}, \texttt{\{liwujun\}@nju.edu.cn}, 
}

\begin{document}

\maketitle

\begin{abstract}

The high memory and computation demand of large language models~(LLMs) makes them challenging to be deployed on consumer devices due to limited GPU memory. 
Offloading can mitigate the memory constraint but often suffers from low GPU utilization, leading to low inference efficiency. 
In this work, we propose a novel framework, called \underline{pip}elined \underline{o}ffloading~(PIPO), for efficient inference on consumer devices. 
PIPO designs a fine-grained offloading pipeline, complemented with optimized data transfer and computation, to achieve high concurrency and efficient scheduling for inference. 
%
Experimental results show that PIPO can outperform existing methods. 
In particular, compared with state-of-the-art baseline, PIPO increases GPU utilization from below 40\% to over 90\% and achieves up to 3.1× higher throughput, running on a laptop equipped with a RTX3060 GPU of 6GB memory.

\end{abstract}

\section{Introduction}

Large language models~(LLMs) have demonstrated strong capabilities in many applications, such as writing, conversation and code generation, in recent years~\cite{OPT,llama3}. 
Due to the increasing demand for privacy preservation~\cite{privacy} and reducing deployment cost~\cite{flexgen}, inference of LLMs on local devices~(typically consumer devices) has been more and more important. 

However, the high memory and computation demand of LLMs makes them difficult to be deployed on consumer devices such as personal computers~(PCs), since these devices typically use GPUs with limited memory. 
For example, the models with 7B or 30B parameters require 15GB to tens of GB of memory for storing model weights, which may be fitted on RTX3090 with 24GB memory but cannot be fitted on other RTX30 series with 6-12GB memory. 
Moreover, the Key-Value~(KV) cache, essential for optimizing attention computations, grows proportionally with sequence length and batch size, further exacerbating the memory insufficiency problem.

To achieve the goal of fitting LLMs into GPUs with limited memory, three techniques have been widely used, including quantization~\cite{awq,1bitllm,smoothquant}, sparsification~\cite{dejavu,powerinfer,minference} and offloading~\cite{deepspeed,llama.cpp}. 
Quantization uses lower precision of data and thus compresses the model. 
Sparsification improves efficiency by pruning and removing some weights and KV-cache. 
Offloading transfers weights and KV-cache from GPU memory to CPU memory or disk, for overcoming GPU memory limit.

Although the quantization methods can decrease the model size by 4$\times$ with 4-bit precision~(assuming the original value is of 16 bits), it remains insufficient for deploying a 30B model after quantization on a GPU with 8GB memory or less. 
Sparsification can greatly decrease the model size by up to 80\%, but it suffers from non-negligible accuracy degradation. 
Hence, offloading has become a necessary strategy for deploying LLMs, especially LLMs with relatively large size, on consumer devices. 

Offloading adopts multi-level memory to schedule the excessive memory demands during the LLMs' inference. 
For example, llama.cpp~\cite{llama.cpp} distributes model layers between the CPU and GPU, leveraging both for inference. 
However, this hybrid inference approach is hindered by CPU's limited computational ability on consumer devices, resulting in high inference latency. 
FlexGen~\cite{flexgen} focuses on GPU-centric computation, utilizing a three-level memory hierarchy that includes GPU memory, CPU memory and disk. 
During LLMs' inference, activation, weight and KV-cache are scheduled to be transferred between these memory levels, overlapping with the computation on the GPU.

However, there still existed two issues in existing offloading frameworks: insufficient inference concurrency and underutilization of disks' bandwidth.

\emph{Insufficient Inference Concurrency:} 
While LLMs are often viewed as memory-bound due to the frequent matrix multiplication operations with low arithmetic intensity, offloading exacerbates this issue, causing high data transfer latency. 
For instance, even with overlap techniques applied in the FlexGen framework, its coarse-grained control and low concurrency result in substantial inefficiency. 
In the case of OPT-30B with CPU-offloading, over 90\% of the inference time is spent on data transfer, with GPU computation accounting for only about 5\%. 
This insufficient inference concurrency leads to significant GPU idle time~(low GPU utilization) and severely limits throughput. 

\emph{Underutilization of Disks' Bandwidth:} 
Although many of the existing offloading frameworks like FlexGen support disk-offloading, they primarily rely on CPU-offloading, requiring up to 200GB of CPU memory to offload the entire model. 
This far exceeds the typical memory capacity of consumer devices and neglects the potential of disks in LLM inference. 
In contrast, NVMe M.2 solid-state Drives~(SSDs) offer significantly higher bandwidth than hard disk drives~(HDDs), providing a more practical solution for offloading on consumer devices.
Unfortunately, existing methods do not effectively utilize SSD bandwidth and pay insufficient attention to disk-based optimizations, leading to inefficient performance in disk-offloading inference.

In this work, we propose a novel framework, called \underline{pip}elined \underline{o}ffloading~(PIPO), for efficient inference on consumer devices by addressing the above two issues.
The main contributions of PIPO are outlined as follows:  

\begin{itemize}

\item
PIPO designs a fine-grained offloading pipeline with high concurrency that optimally balances memory usage and inference efficiency, supporting larger models on limited GPU memory. 

\item
PIPO emphasizes the crucial role of NVMe SSDs in offloading on consumer devices and designs a comprehensive suite to optimize data transfer speed between GPU memory, CPU memory and disk. 

\item
%
Experimental results show that PIPO can outperform existing methods. In particular, compared with state-of-the-art baseline, PIPO increases GPU utilization from below 40\% to over 90\% and achieves up to 3.1$\times$ higher throughput, running on a laptop equipped with a RTX3060 GPU of 6GB memory. 

\end{itemize}

\section{Related Work}

With the growing interest in deploying LLMs on local devices, inference frameworks based on offloading strategies have gained significant attention and research focus. 
\mbox{DeepSpeed}~\cite{deepspeed} and HuggingFace Accelerate~\cite{huggingface} are among the first systems to integrate offloading into inference. 
They directly adopt offloading techniques from training, without in-depth study of the distinct characteristics of the inference phase. 
\mbox{FlexGen}~\cite{flexgen} proposes an efficient offloading strategy through a search-based algorithm, optimizing data placement and overlapping computations with I/O operations. 
Powerinfer~\cite{powerinfer} and LLM in a flash~\cite{llminaflash} both utilize sparsification by keeping hot neurons on the GPU while offloading the rest weights to the CPU. 
During inference, they dynamically fetch necessary neurons from storage to device's memory for computation. 
Although this improves the inference speed, it still suffers from accuracy degradation and high CPU memory overhead. 
While these offloading-based frameworks explore offloading strategies, they fall short of a thorough analysis of the associated performance overhead. 
FlexGen attempts to enhance efficiency through overlapping, but it still falls short of fully leveraging concurrency in the inference process. 
In addition to the strategy of loading data to the GPU for computation during inference, an alternative approach is to distribute computations across both the GPU and CPU. 
llama.cpp~\cite{llama.cpp} implements this by partitioning the model at the Transformer layer level, storing some weights on the GPU while offloading the rest to the CPU. 
The CPU processes its assigned layers first and then transfers the results to the GPU for further computation. 
While this method reduces PCIe data transfer, it results in substantial GPU idle time and slows down inference. 

From an algorithmic perspective, quantization~\cite{smoothquant,awq,omniquant,qserve,squeezellm,vptq,1bitllm} has been widely adopted for LLM inference to reduce the memory footprint and accelerate the computation. 
For weight-only quantization, recent research has achieved extremely low-bit precision, down to 3 bits and even below 2 bits~\cite{squeezellm,vptq,1bitllm}. 
However, such aggressive quantization faces significant challenges in practical applications, including deployment complexity and potential degradation in inference speed. 
In contrast, simultaneously quantizing weights, activations, and KV-cache to relatively low-bit precisions (e.g., W8A8, W4A4, and W4A8KV4) has been shown to balance memory efficiency and inference throughput~\cite{smoothquant, omniquant, qserve}. 
PIPO supports quantizing both weights and KV-cache to INT4 to accommodate the limited memory of consumer devices and enhance inference throughput. 

\begin{figure}
  \centering
  \includegraphics[width=\textwidth]{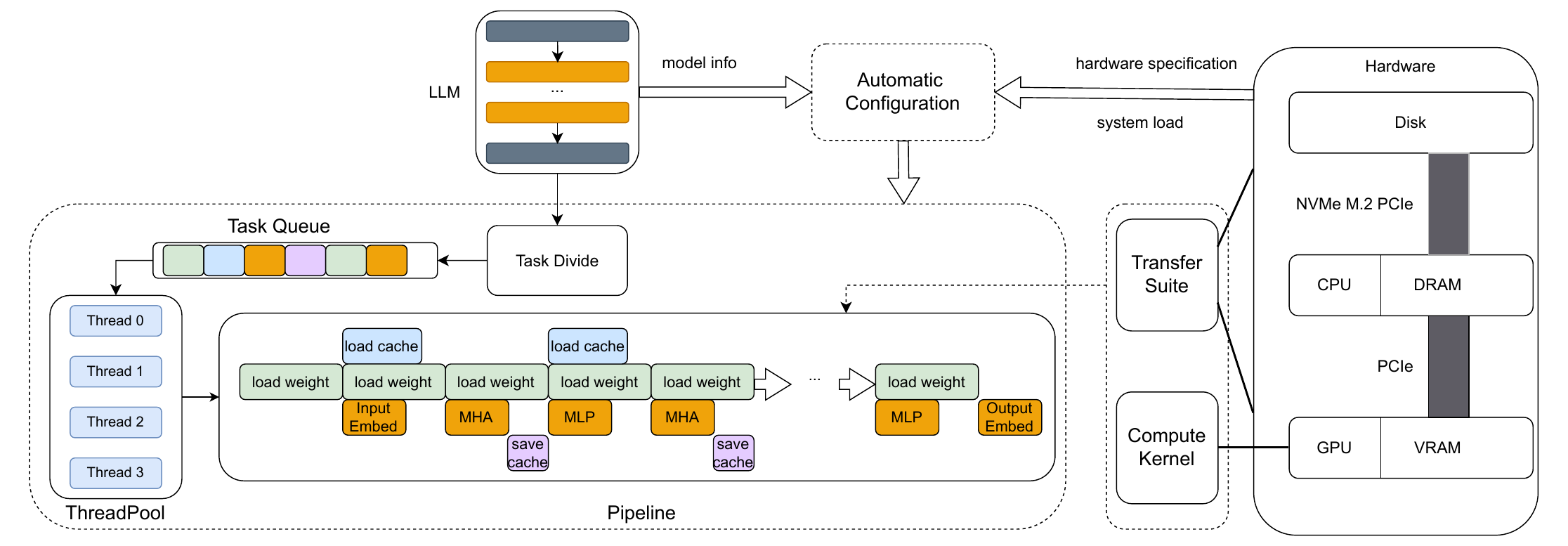}
  \caption{The architecture of PIPO.}
  \label{overview}
\end{figure}

\section{Pipelined Offloading}
Pipelined offloading~(PIPO) is an offloading inference framework with high throughput designed for consumer devices with limited GPU and CPU memory. 
As shown in Figure~\ref{overview}, the architecture of PIPO consists of three key components: pipeline and thread pool, transfer suite and compute kernel optimizations, automatic configuration

\emph{Pipeline and Thread Pool:}  
A fine-grained inference pipeline with an efficient thread pool maximizes concurrency in offloading inference, enhancing hardware utilization and inference throughput. 

\emph{Transfer Suite and Compute Kernel Optimizations:} 
A comprehensive suite optimizes data transfer between disk, CPU memory, and GPU memory. Furthermore, custom quantized compute kernels minimize computation latency. 

\emph{Automatic Configuration:} 
PIPO automatically determines the most efficient offloading, pipeline strategies, and resource allocation before execution, based on hardware specifications, model parameters, and system load.

PIPO starts by analyzing key input parameters, including model details~(e.g., model size, batch size, and precision), hardware specifications, and the current system workload. 
Leveraging this information, PIPO automatically configures the offloading and pipeline strategy, while applying optimization techniques, including those for data transfer and computation, to maximize overall performance. 
During inference, the workload is divided into four distinct task types: computation, weight loading, KV-cache loading, and KV-cache saving, which are sequentially organized into a task queue. 
A thread pool retrieves tasks from the queue and executes them in parallel, while the main thread coordinates synchronization between tasks to ensure fine-grained control and minimize unnecessary delays. 

The following content of this section will present the details about PIPO, including offloading architecture, pipeline design, data transfer suite, computation optimization, and automatic configuration. 

\subsection{Offloading Architecture} 
In this subsection, we introduce the offloading architecture, including offloading strategies and task design. 

\subsubsection{Offloading Strategies}
PIPO stores the entire model's weights in GPU memory, CPU memory, or NVMe disk, depending on available capacity.
The KV-cache is stored in CPU memory, with the GPU loading the required cache before computing the multi-head attention~(MHA) layer and storing the newly generated cache back to CPU memory after computation. 
After computing each layer, the outputs are retained in GPU memory as inputs for the subsequent layer, while the memory occupied by the weight and cache is released to accommodate the next layer. 
This strategy significantly reduces GPU memory usage during inference by storing only one or a few layers~(in the case of preloading) on the GPU at a time. 
Existing state-of-the-art~(SOTA) models like LLaMA3.1 contain 62 and 162 layers in the 8B and 70B models, respectively, when treating MHA and MLP as separate layers. 
By loading only a small fraction of the model at once, PIPO drastically reduces memory usage, enabling the majority of GPU memory to be allocated for supporting long context lengths or large batch sizes.

An alternative approach to PIPO is to store as many of the weights in GPU memory as possible and load the remainder from CPU memory and disk. 
The benefit of this approach is that it eliminates the latency of loading weights already stored on the GPU. 
However, it has two key drawbacks: 
(i) latency reduction is minimal, as only a small portion of the model weights can be stored in GPU memory, 
and (ii) even when a larger portion of the model can be stored in GPU memory, the available GPU memory can only accommodate small batch sizes, significantly limiting inference throughput. 

\subsubsection{Task Design}

For higher inference performance and resource management, PIPO adapts and modifies existing task division approaches~\cite{flexgen}, dividing inference into four task types: computation, weight loading, KV-cache loading, and KV-cache saving.

\emph{Computation} 
encompasses MHA and MLP layers, as well as the input and output embedding layers at the beginning and end of the model. 
Due to the sequential nature, all computation tasks must be executed one by one. 

\emph{Weight Loading} 
involves transferring model weights from CPU memory or disk to GPU memory.
As weight loading is often a bottleneck due to PCIe bandwidth limitations, it needs to be executed concurrently with computation to improve efficiency.
However, the latency of weight loading is typically longer than computation. 
Loading weights too early may increase memory usage without improving performance. 
To address this, the weight loading for each layer begins only after the previous loading task is completed, overlapping with the computation of the previous layer.

\emph{KV-cache Loading} 
transfers the necessary KV-cache from CPU memory to GPU memory for token generation. 
Similarly, PIPO schedules KV-cache loading to overlap with computation from the preceding layer. 
Since KV-cache is only used in MHA layers, its execution can be optimized by advancing it one layer ahead to overlap with computation from previous MHA layer, further enhancing pipeline concurrency. 

\emph{KV-cache Saving} 
stores newly generated KV-pairs back to CPU memory for future token generation. 
Unlike loading tasks, KV-cache saving provides greater flexibility in synchronization. 
It is scheduled immediately after MHA computation, but its completion is only ensured when the saved cache is required for the current layer in the next token generation cycle.

PIPO decouples preprocessing and postprocessing of weights and KV-cache from computation and assigns them to loading tasks, as their sequential dependencies with data transfer prevent concurrent execution. 

\subsection{Pipeline Design}
\label{pipeline_design}

A well-designed pipeline is essential for achieving high concurrency during inference while minimizing resource overhead.
The pipelined inference is backed by a thread pool, which provides the foundation for concurrent execution, enabling efficient task coordination within the pipeline.
In this section, we detail the design of the pipeline, focusing on thread pool configuration, pipeline scheduling, and the performance-memory tradeoff.

\subsubsection{Thread Pool Configuration}
During inference, the thread pool retrieves tasks from the task queue and assigns them to individual threads for execution. 
Keeping the thread pool compact and tidy is essential for maintaining simplicity and avoiding unnecessary overhead. 
We design the thread pool based on several principles. 
Except for KV-cache saving, only one instance of each operation type can execute at a time due to the sequential nature of LLM inference. 
KV-cache saving supports executing multiple requests and has relatively delayed synchronization. 
It requires minimal computational and bandwidth resources and is assigned a lower priority in the scheduling hierarchy. 
To reduce idle time caused by waiting for other threads, the main thread not only schedules tasks within the pipeline but also directly handles computation. 

Consequently, PIPO configures the thread pool size to three, corresponding to data transfer types, while computation is handled outside the pool by the main thread. 
Crucially, the threads within the pool are not statically assigned to specific tasks. 
Since execution time may vary across layers, this flexible scheduling approach allows threads to dynamically handle incoming tasks, minimizing idle time.

\begin{figure}
  \centering
  \includegraphics[width=\textwidth]{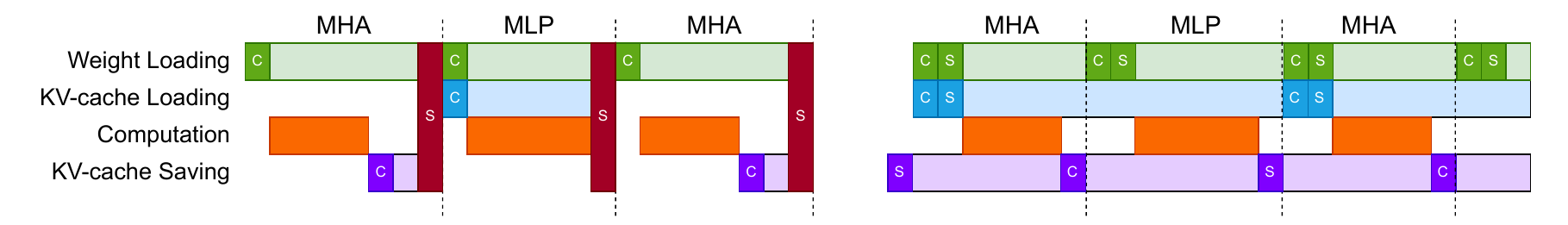}
  \caption{Comparison of pipeline designs in FlexGen~(left) and PIPO~(right). `C' denotes `Call' which instructs the device to execute a given task, while `S' denotes `Synchronize' to ensure the completion of a single task~(or synchronizing the device in FlexGen).}
  \label{pipeline}
\end{figure}

\begin{algorithm}[t]
   \caption{Pipeline Scheduling}
   \label{alg}
\begin{algorithmic}
   \FOR{$i \in \text{generation\_length}$}
       \FOR{$j \in \text{num\_layer}$}
           \STATE CallLoadData($i, j$); {\footnotesize \texttt{ \# Preload weight and cache for subsequent layers}}
           \STATE PrepareInput($i, j$); {\footnotesize \texttt{ \# Prepare hidden and mask}}
           \STATE SynchronizeLoadTask($i, j$); {\footnotesize \texttt{ \# Synchronize current layer's data loading tasks}}
           \STATE Compute($i, j$); {\footnotesize \texttt{ \# Current layer's computation}}
           \IF{$\text{layer[j]} = \text{MHA}$}
               \STATE CallStoreCache($i, j$); {\footnotesize \texttt{ \# Store KV-cache before loaded}}
           \ENDIF
       \ENDFOR
   \ENDFOR
\end{algorithmic}
\end{algorithm}

\subsubsection{Pipeline Scheduling}

As illustrated in Figure~\ref{pipeline}, PIPO's pipeline design significantly enhances pipelining efficiency and reduces idle time~(bubbles), compared to FlexGen's overlapping approach. 
By implementing task-level synchronization, PIPO achieves precise control over task execution, minimizing unnecessary delays between tasks.
Furthermore, PIPO adjusts the synchronization order and timing for each task, unlocking greater efficiency and improving GPU utilization.

The pipeline scheduling process is detailed in Algorithm~\ref{alg}. 
During LLM inference for a specific layer, PIPO first initiates the loading of weights and KV-cache~(if the next layer involves an MHA operation) for the next layer. 
Simultaneously, it prepares the input data for the current layer's computation and synchronizes the data transfer tasks for the current layer to ensure the necessary data is available for computation.
Computation is executed on the main thread, with other threads in the pool concurrently handling the data transfer tasks in the queue. 
After computation completes, PIPO first launches the KV-cache saving task~(if the current layer is MHA), and stores the computation output. 
A key aspect of the pipeline is ensuring KV-cache saving finishes before the same layer's KV-cache loading in the next token generation loop. 
If saving tasks are delayed, the required cache cannot be correctly loaded. 
To address this, PIPO advances the completion check for KV-cache saving to one layer earlier, ensuring the required cache is ready when needed for subsequent loading and computation.  

\subsubsection{Performance-Memory Tradeoff}
PIPO's pipeline achieves high performance but incurs certain memory overhead.  
For higher throughput, PIPO preloads the weight and KV-cache for the next layer, storing data of two layers in GPU memory simultaneously. 
Additionally, PIPO synchronizes KV-cache saving tasks to complete before the same layer in the next loop, which could temporarily store all generated KV-pairs on the GPU in extreme cases.
During decoding, the amount of KV-pairs is equal to the number of MHA layers. 
This size is significantly smaller than that of the KV-cache loaded for a single layer, which is equal to the input length. 
When decoding LlamA3.1-8B, the GPU memory usage remains under 2GB, which is feasible on almost all the consumer GPUs.

However, the memory demand is significantly higher during the prefill stage. 
This stage processes the entire input sequence and generates all the KV-cache for the sentence at once, leading to substantial memory usage. 
To mitigate this, PIPO reduces task concurrency but lowers memory usage, ensuring the model can be fitted within available GPU memory. 
For instance, by synchronizing the cache saving task before launching the next one, only a single KV-pair occupies GPU memory at a time. 
At the lowest memory usage, PIPO only requires the weights and KV-cache of a single model layer to perform inference.
In conclusion, PIPO offers two pipeline options: a \emph{performance-optimized pipeline} and a \emph{memory-efficient pipeline}. In real applications, one of these two options will be automatically chosen based on the available resources, which will be detailed in Section~\ref{sec:dynamicConfiguration}.

\begin{figure}
    \centering
    \begin{minipage}[b]{0.48\textwidth}
        \centering
        \includegraphics[width=\linewidth]{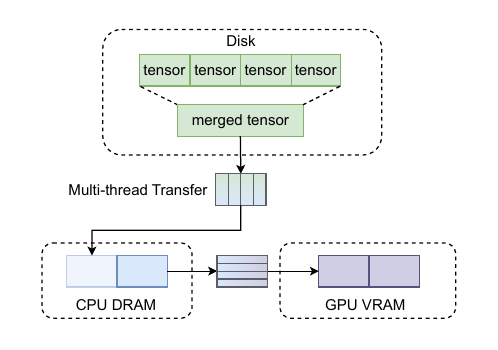}
        \caption{Blockwise data transfer timeline.}
        \label{timeline}
    \end{minipage}
    \begin{minipage}[b]{0.48\textwidth}
        \centering
        \includegraphics[width=\linewidth]{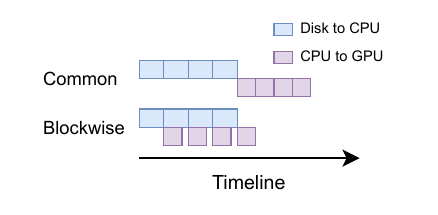}
        \caption{Disk to GPU memory transfer.}
        \label{suite}
    \end{minipage}
\end{figure}



\subsection{Data Transfer Suite} 
To address the data transfer challenges introduced by offloading, PIPO incorporates a specialized data transfer suite, replacing standard PyTorch and NumPy methods.
These challenges arise from the need to transfer large volumes of data across memory hierarchies, particularly from disk to GPU memory.
The suite optimizes critical data movement and improves bandwidth utilization, enhancing transfer speed. 
As illustrated in Figure~\ref{suite}, the suite employs several techniques to enhance disk to GPU memory transfer efficiency in disk-offloading, including blockwise transfer, muti-thread parallel transfer, and data merging.

\emph{Blockwise Transfer:}
PIPO divides weight tensors into manageable blocks, transferring them in a pipelined manner across stages: from disk to CPU memory and from CPU memory to GPU memory. 
For example, while one block is being read into CPU memory, another block can be simultaneously transferred to GPU memory. 
This overlapping reduces idle time and boosts overall effective bandwidth, as illustrated in the timeline in Figure~\ref{timeline}. 

\emph{Multi-thread Parallel Transfer:}
With each block further divided into smaller chunks, PIPO employs multiple CPU threads to handle loading chunks from disk, while GPU threads manage their transfer to GPU memory.
As soon as a CPU thread finishes its chunk, it signals the GPU thread to proceed, maintaining a continuous data flow and reducing idle time. 

\emph{Data Merging:} 
During LLM inference, weight tensors within the same layer are loaded separately.
To reduce the overhead caused by frequent I/O requests, PIPO merges these weight tensors into a single tensor.
This merged tensor can be loaded with a single request and further divided into multiple blocks for blockwise transfer, enabling higher parallelism and improved effective bandwidth. 

In addition to optimizing disk to GPU memory transfer, PIPO also leverages multi-thread parallel transfer and data merging techniques for transfers between CPU and GPU memory. 
These optimizations effectively manage data movement across each transfer stage, enhancing overall data transfer performance.

\subsection{Computation Optimization}  
Although data transfer is the primary bottleneck in offloading inference, computation also requires careful optimization, particularly with quantized weights.  
In conventional approaches, quantized weights are dequantized into floating-point values~(e.g., half-precision) before computation, incurring both time and memory overhead.  
To address this problem, PIPO introduces custom handwritten compute kernels that perform matrix-vector multiplication directly on 4-bit quantized weights, avoiding the dequantization operation.
The kernels are highly optimized to leverage the full computational power of the GPU to improve performance, particularly in scenarios with small batch sizes where the GPU is typically underutilized. 

\subsection{Automatic Configuration} \label{sec:dynamicConfiguration} 

PIPO automatically configures~(determines) the optimal offloading and pipeline strategies based on user-specified parameters, hardware specifications, and system load.
It considers factors such as model type, precision, batch size, prompt and generation lengths, GPU and CPU memory, PCIe bandwidth, and current resource utilization to automatically adjust and optimize the inference process.
Using these inputs, PIPO determines the offloading strategy by assigning weights to GPU memory, CPU memory, or disk based on memory constraints. 
It also selects the pipeline strategy, choosing between a performance-optimized pipeline and a memory-efficient pipeline to balance inference performance and memory usage.

Taking LLaMA3.1 model family as an example, we outline the configuration process for PIPO. 
Assume the model has \( l \) hidden layers, an input dimension of \( d \), and a vocabulary size of \( V \). 
Additionally, \( p \) represents the data type size~(precision), \( b \) is the batch size, and \( s \) denotes the input length which is the sum of the prompt length and the generated length. 
The LLaMA3.1 models employ Grouped-Query Attention~(GQA), with the number of attention heads denoted by \( h \) and the number of KV heads denoted by \( h_{\text{kv}} \).  
%
The hidden dimension \( d_{\text{h}} \) is calculated as: 
\(d_{\text{h}} = m \cdot \left\lceil \gamma \cdot \left\lfloor \frac{8}{3}d \right\rfloor / m \right\rceil\), where \( m \) and \( \gamma \) are constant values.

The total model weight size is calculated as 
\( W = 2 W_{\text{embed}} + l \cdot (W_{\text{mha}} + W_{\text{mlp}}) \), where 
\( W_{\text{embed}} = pdV \), 
\( W_{\text{mha}} = pd \cdot (2d + d\frac{h_{\text{kv}}}{h} + 1) \), and 
\( W_{\text{mlp}} = pd \cdot (3 \cdot d_{\text{h}} + 1) \).
The total KV-cache size is calculated as 
\( C = 2 \cdot pbsld \cdot \frac{h_{\text{kv}}}{h} \). 
For peak memory usage, PIPO considers the prefill stage with preloading which demands significantly more memory than the decoding stage. 
When employing preloading, the peak memory usage is  
\( M = \max(M_{\text{mha}}, M_{\text{mlp}}, M_{\text{embed}}) \), where
\( M_{\text{mha}} = pbs(5d+hs) + W_{\text{mha}} + W_{\text{mlp}} + 2\frac{C}{l} \), 
\( M_{\text{mlp}} = pbs(3d_{\text{h}}+2d) + W_{\text{mha}} + W_{\text{mlp}} + \frac{C}{l} \), and 
\( M_{\text{embed}} = pbs \cdot (V + d) + 2W_{\text{embed}} \), as detailed in Appendix \ref{Appendix:memory-constrains}. 

To get the optimal memory hierarchy for offloading weights and determining the pipeline strategy, PIPO gathers additional system and hardware information, including the available GPU memory \( M_{\text{GPU}} \), CPU memory \( M_{\text{CPU}} \), GPU PCIe bandwidth \( B_{\text{GPU}} \), and disk PCIe bandiwith \( B_{\text{SSD}} \). 
Then with the given parameters, the configuration of PIPO can be automatically determined through the process outlined in Eq.~(\ref{strategy}).

\begin{small}
\begin{align}
\label{strategy}
\text{Weight on:} 
& \begin{cases} 
  \text{GPU,} & \text{if } W + M < M_{\text{GPU}} \\
  \text{CPU,} & \text{if } W + C < M_{\text{CPU}} \text{ and } B_{\text{SSD}} < B_{\text{GPU}} \\
  \text{Disk,} & \text{else.}
\end{cases} \nonumber \\ 
\text{Pipeline:} 
& \begin{cases} 
  \text{Performance-optimized,} & \text{if } M < M_{\text{GPU}} \\
  \text{Memory-efficient,} & \text{else.}
\end{cases}
\end{align}
\end{small}

Once the offloading and pipeline strategies are determined, PIPO automaticaly adjusts parameters for optimal performance.
This involves enabling \emph{Transfer Suite} for offloading weights, determining the \emph{Block Size} through experiments~(detailed in Appendix~\ref{Appendix:block-size}), and activating \emph{Compute Kernel} for INT4 weights to bypass dequantization overhead with batch sizes less than 16. 

With its automatic configuration capabilities, PIPO delivers a fully automated and highly efficient inference workflow, seamlessly adapting to diverse hardware specifications and varying workloads with minimal manual intervention. 
This flexibility ensures high inference throughput across a wide range of scenarios, reducing the need for user-driven adjustments at the same time. 
The overall workflow of PIPO is succinctly outlined in Algorithm~\ref{alg:workflow}. 

\begin{algorithm}[t]
   \caption{PIPO Workflow}
   \label{alg:workflow}
\begin{algorithmic}
\STATE {\bfseries Input:} model $\mathcal{M}$, batch size $b$, length $s$, precision $p$, CPU memory $M_{\text{CPU}}$, \\
\hspace*{1cm} GPU memory $M_{\text{GPU}}$, GPU bandwidth $B_{\text{GPU}}$, SSD bandwidth $B_{\text{SSD}}$

\STATE $S_{\text{off}}, S_{\text{pipe}} =$ Configure($\mathcal{M}, b, s, p, M_{\text{CPU}}, M_{\text{GPU}}, B_{\text{GPU}}, B_{\text{SSD}}$); {\footnotesize \texttt{ \# Automatic Configuration}}
\STATE InitModel($\mathcal{M}, b, s, p$); {\footnotesize \texttt{ \# Init Model Data}}
\STATE InitTransferSuitAndOperators($\mathcal{M}, b, p, S_{\text{off}}$); {\footnotesize \texttt{ \# Init PIPO components}}
\STATE ConstuctTaskandQueue($\mathcal{M}$); {\footnotesize \texttt{\# Build Inference Runtime}}
\STATE PipelineScheduling( ); {\footnotesize \texttt{\# Generation (Call Algorithm~\ref{alg})}}
\end{algorithmic}
\end{algorithm}

\section{Experiment}
\subsection{Experimental Setting}
{\emph{Hardware.}} 
We conduct experiments on a Lenovo Thinkbook and a desktop\footnote{Due to space limitation, the results for the desktop are moved to Appendix~\ref{Appendix:high-end-performance}. Furthermore, 
Appendix~\ref{Appendix:parallelism} provides a discussion on PIPO's combination with parallelism techniques for multi-GPU settings. 
}. The Lenovo Thinkbook is a typical consumer laptop which is equipped with NVIDIA RTX3060~(6G), 16GB CPU Memory and 1TB M.2 SSD.

{\emph{Model.}}
We evaluate PIPO on three LLM families: OPT~(6.7B, 13B, 30B and 66B) for benchmarking a wide range of model sizes, LLaMA3.1~(8B and 70B) for assessing the performance on the latest models, and MoE models\footnote{Due to space limitation, the results for MoE models,  including Mixtral 8×7B~\cite{mixtral} and DeepSeek-R1 671B~\cite{deepseek}, are moved to Appendix~\ref{Appendix:moe}.}. 

{\emph{Workload.}}
To simulate real-world inference tasks, we evaluate PIPO on text generation with varying sequence lengths and batch sizes. 
The prompt length is set to 512 tokens, and PIPO generates 32 tokens for each prompt\footnote{More experiments on varying context lengths~(512-3072 tokens) are  provided in Appendix~\ref{Appendix:context-length}, which demonstrates stable inference performance of PIPO.}. 
The batch size ranges from 1 to 32.
In our experiments, all models use FP16~(or BF16) and INT4 precision for weights, with intermediate activations in FP16~(or BF16), which is consistent with recent LLM research practices. 

{\emph{Implementation.}}
PIPO is built upon a reconstruction and extension of FlexGen by retaining critical data structures in FlexGen, while incorporating new modules in C++/CUDA and Python code to improve efficiency and offers excellent extensibility. 
%

{\emph{Baseline.}}
FlexGen has achieved state-of-the-art performance~\cite{flexgen} in offloading inference. Hence, we choose FlexGen as baseline. 
For a fair comparison, we use the consistent weight storage type for both FlexGen and PIPO and extend FlexGen to support LLaMa models, as FlexGen does not natively support them. 
All experimental results are averaged over at least 3 independent runs.

\begin{figure}
  \centering
  \includegraphics[width=0.9\textwidth]{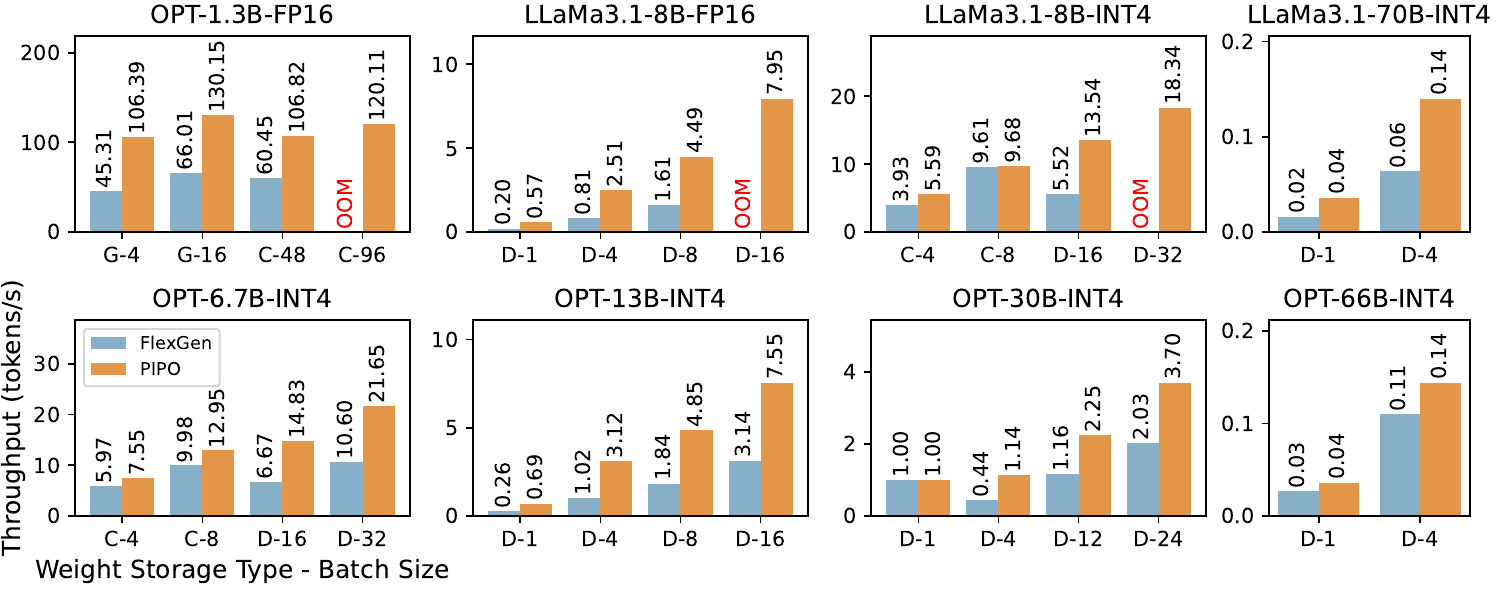}
  \caption{Throughput of models.
     The X axis indicates the weight's storage type and batch size. For example, `G-4', `C-8' and `D-16' indicate weight on GPU with batch size of 4, CPU-offloading with batch size of 8 and disk-offloading with batch size of 16, respectively.}
  \label{g}
\end{figure}

\subsection{Results}
We first evaluate the end-to-end inference throughput on LLM models with varied model size and data precision that PIPO can achieve, as shown in Figure~\ref{g}. 
%
%
When model size or batch size increases, the activation in GPU memory and the KV-cache in CPU memory increase proportionally.
Hence, PIPO offloads weights to CPU memory and disk when GPU or CPU memory is insufficient. 
For the OPT-1.3B model, which is commonly deployed on consumer devices, PIPO achieves an average throughput improvement of 2.03$\times$.
For larger models like OPT-6.7B~(about 13GB of weights) and LLaMA3.1-8B~(more than 16GB of weights), disk-offloading becomes necessary. 
PIPO achieves up to 3.10$\times$ higher throughput in disk-offloading and outperforms FlexGen in CPU-offloading as well.
%
We conduct a series of experiments on the OPT and LLaMA3.1 model families in INT4 format as well where PIPO outperforms FlexGen in all cases, achieving an average improvement of 1.97$\times$ and a peak improvement of 3.04$\times$.
These results demonstrate the efficiency of PIPO’s pipelined offloading inference architecture and highlight its potential for optimizing LLM inference on consumer devices.

We further investigate the performance improvement provided by PIPO by measuring the transfer speed and GPU utilization. 
PIPO achieves a 26\% improvement in disk-to-GPU transfer speed and enhances GPU utilization from 36\% to 97\%, as shown in Appendix~\ref{Appendix:gpu-utilization}.
Ablation studies confirm PIPO's pipeline scheduling contributes the most significant performance gain~(1.97$\times$ speedup, see Appendix~\ref{Appendix:ablation}).
As demonstrated in Appendix~\ref{Appendix:latency}, PIPO achieves a 42.5\% reduction in time-to-first-token~(TTFT), which is a crucial aspect of inference efficiency for end users as well. 
Finally, we discuss in detail PIPO's memory footprint and offloading's overhead during inference in Appendix~\ref{Appendix:memory} and ~\ref{Appendix:overhead}, which shows that compared to a non-offload implementation, PIPO can reduce VRAM usage by 66.4\% with only 11.2\% performance degradation.



\section{Conclusion}
In this work, we propose a novel offloading inference framework, called PIPO, for efficient LLM inference on consumer devices. 
By leveraging a fine-grained inference pipeline coupled with optimizations in data transfer and computation, PIPO significantly improves GPU utilization and inference throughput.  
Experimental results show that PIPO achieves a remarkable throughput improvement  compared to the widely adopted offloading framework FlexGen. 
PIPO’s flexible architecture ensures its adaptability to various LLMs, making it a promising solution for running LLMs locally on consumer devices such as personal desktops and laptops.

\bibliography{pipo}
\bibliographystyle{unsrt} 

\newpage
\appendix

\section{Experiments about Block Size }
\label{Appendix:block-size}
\begin{figure}[h]
  \centering
  \includegraphics[width=\textwidth]{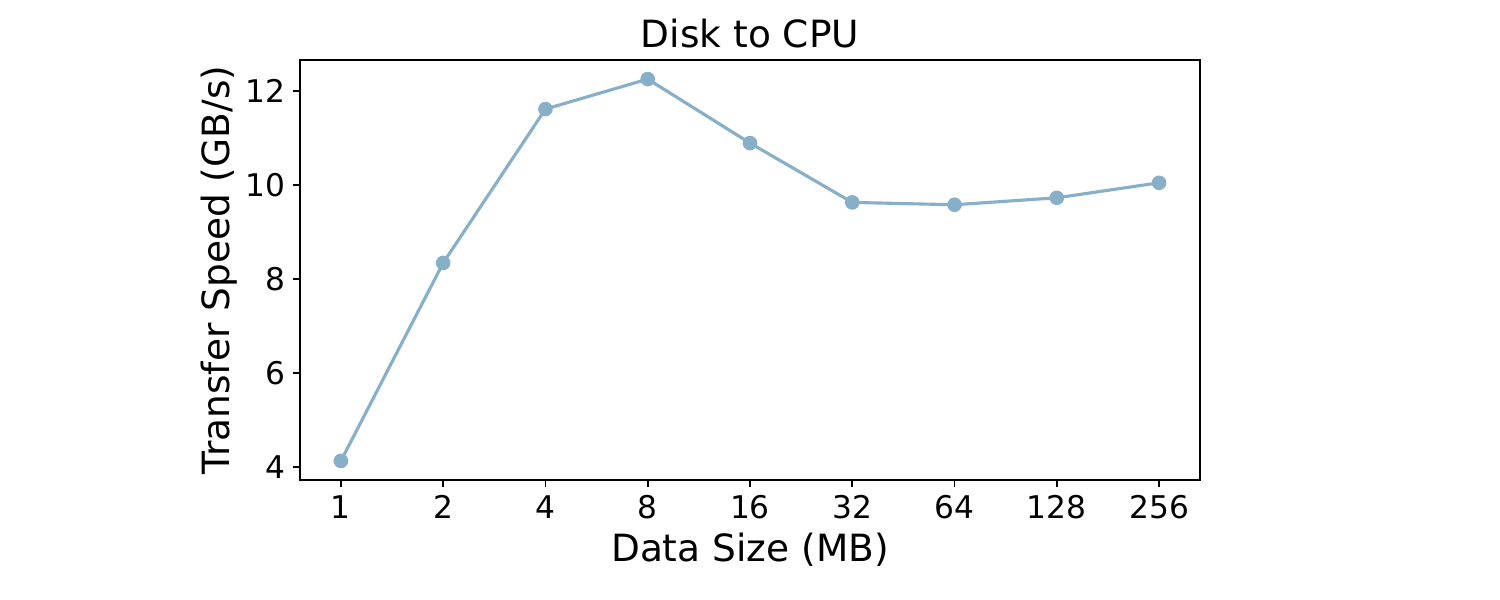}
  \includegraphics[width=\textwidth]{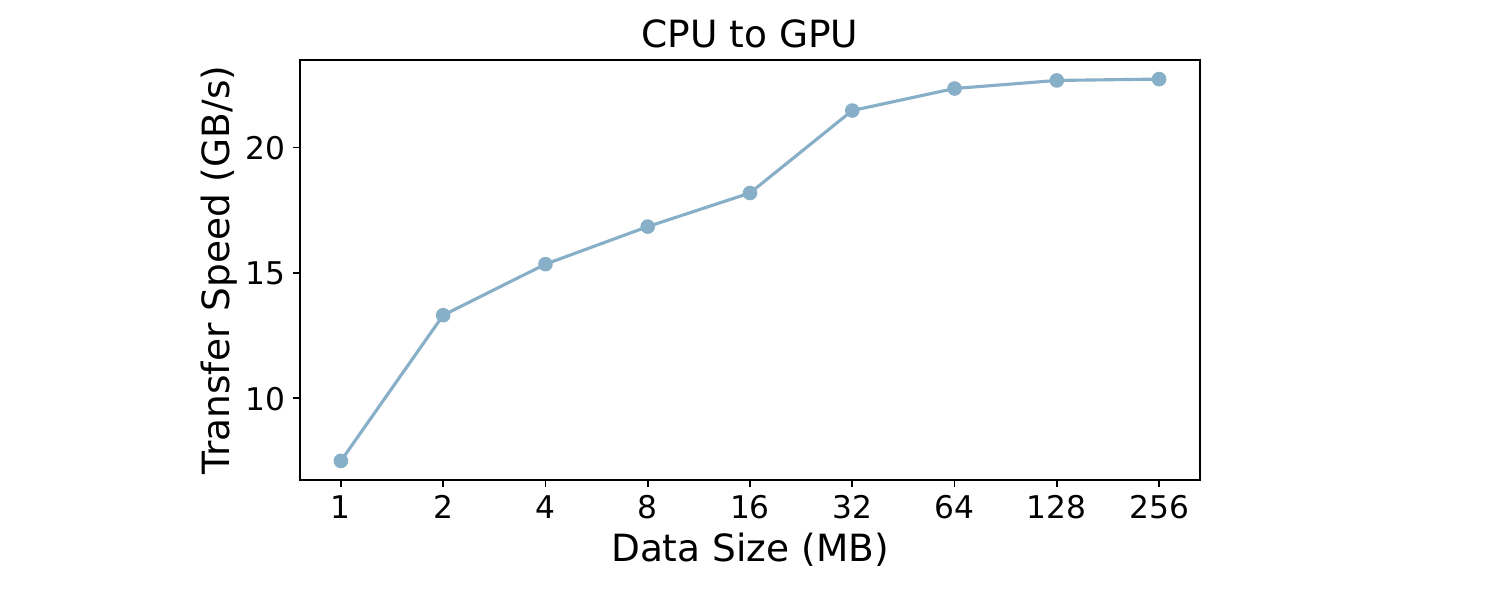}
  \caption{Transfer speed for different data sizes.}
  \label{block}
\end{figure}

Modern storage systems are designed to read and write data in discrete units, such as pages or blocks.
Hardware is typically optimized for high-speed I/O operations on contiguous memory addresses, which significantly improves performance for such access patterns.
Consequently, when reading a complete tensor in blocks, performance is generally not adversely affected as long as the block size is appropriately selected. 
To determine the optimal block size for blockwise transfer, PIPO conducts experiments to measure transfer speeds across different data sizes on the target device.
Example experimental results on a Lenovo Thinkbook equipped with an RTX3060 are shown in Figure~\ref{block}.

For disk to CPU memory transfers, the block size of 8MB achieves the best performance.
For CPU memory to GPU memory transfers, the block sizes above 32MB reach the bandwidth limitation, indicating optimal performance at this threshold.
Hence, PIPO adopts block size of 32MB for overall data transfer on the target device.

\section{Memory Constrains}
\label{Appendix:memory-constrains}
In this section, we demonstrate the memory constrains of LLaMA3.1 with and without preloading for prefill and decoding stages.

\subsection{Prefill Stage}

\begin{align*}
M_{\text{input}} &= pbsd \\
M_{\text{output}} &= pbsd \\
M_{\text{qkv}} &= 3pbsd \\
M_{\text{w}} &= pbsd_{\text{h}} \\
M_{\text{attn}} &= pbhs^2
\end{align*}

With preloading:

\begin{align*}
M_{\text{mha}} &= M_{\text{input}} + M_{\text{qkv}} + M_{\text{attn}} + M_{\text{output}}\\
&\quad+ W_{\text{mha}} + W_{\text{mlp}} + \frac{C}{l} \\
&= pbs(5d+hs) + W_{\text{mha}} + W_{\text{mlp}} + \frac{C}{l} \\
&= pbs(5d+hs) + pd(2d+2d\frac{h_{\text{kv}}}{h}+1) \\
&\quad+ pd(3d_{\text{h}}+1) + 2pbsd\frac{h_{\text{kv}}}{h} \\
M_{\text{mlp}} &= M_{\text{input}} + 3M_{\text{w}} + M_{\text{output}} + W_{\text{mlp}} + W_{\text{mha}} \\
&= pbs(3d_{\text{h}}+2d) + pd(3d_{\text{h}}+1) \\
&\quad+ pd(2d+2d\frac{h_{\text{kv}}}{h}+1) \\
M_{\text{embed}} &= M_{\text{input}} + pbsV + \max(W_{\text{mha}}, W_{\text{embed}}) + W_{\text{embed}} \\
&= pbs(d+V) + \max(pd(2d+2d\frac{h_{\text{kv}}}{h}+1),pdV)\\
&\quad+ pdV \\
&= pbs(d+V) + pdV + pdV
\end{align*}

Without preloading:

\begin{align*}
M_{\text{mha}} &= M_{\text{input}} + M_{\text{qkv}} + M_{\text{attn}} + M_{\text{output}}\\
&\quad+ W_{\text{mha}} + \frac{C}{l} \\
&= pbs(5d+hs) + pd(2d+2d\frac{h_{\text{kv}}}{h}+1) \\
&\quad+ 2pbsd\frac{h_{\text{kv}}}{h} \\
M_{\text{mlp}} &= M_{\text{input}} + 3M_{\text{w}} + M_{\text{output}} + W_{\text{mlp}}\\
&= pbs(3d_{\text{h}}+2d) + pd(3d_{\text{h}}+1) \\
M_{\text{embed}} &= M_{\text{input}} + pbsV + W_{\text{embed}} \\
&= pbs(d+V) + pdV
\end{align*}

\subsection{Decoding Stage}
The decoding stage has an input length of 1, which significantly decreases the memory requirement in generation.

\begin{align*}
M_{\text{input}} &= pbd \\
M_{\text{output}} &= pbd \\
M_{\text{qkv}} &= 3pbd \\
M_{\text{w}} &= pbd_{\text{h}} \\
M_{\text{attn}} &= pbh
\end{align*}

With preloading:

\begin{align*}
M_{\text{mha}} &= M_{\text{input}} + M_{\text{qkv}} + M_{\text{attn}} + M_{\text{output}}\\
&\quad+ W_{\text{mha}} + W_{\text{mlp}} + \frac{2C}{l} \\
&= pb(5d+h) + W_{\text{mha}} + W_{\text{mlp}} + \frac{2C}{l} \\
&= pb(5d+h) + pd(2d+2d\frac{h_{\text{kv}}}{h}+1) \\
&\quad+ pd(3d_{\text{h}}+1) + 4pbsd\frac{h_{\text{kv}}}{h} \\
M_{\text{mlp}} &= M_{\text{input}} + 3M_{\text{w}} + M_{\text{output}} + W_{\text{mlp}} + W_{\text{mha}} + \frac{C}{l} \\
&= pb(3d_{\text{h}}+2d) + pd(3d_{\text{h}}+1) \\
&\quad+ pd(2d+2d\frac{h_{\text{kv}}}{h}+1) + 4pbsd\frac{h_{\text{kv}}}{h} \\
M_{\text{embed}} &= M_{\text{input}} + pbsV + \max(W_{\text{mha}}, W_{\text{embed}}) + W_{\text{embed}} \\
&= pb(d+V) + \max(pd(2d+2d\frac{h_{\text{kv}}}{h}+1),pdV)\\
&\quad+ pdV \\
&= pb(d+V) + pdV + pdV
\end{align*}

Without preloading:

\begin{align*}
M_{\text{mha}} &= M_{\text{input}} + M_{\text{qkv}} + M_{\text{attn}} + M_{\text{output}}\\
&\quad+ W_{\text{mha}} + \frac{C}{l} \\
&= pb(5d+hs) + pd(2d+2d\frac{h_{\text{kv}}}{h}+1) \\
&\quad+ 2pbsd\frac{h_{\text{kv}}}{h} \\
M_{\text{mlp}} &= M_{\text{input}} + 3M_{\text{w}} + M_{\text{output}} + W_{\text{mlp}}\\
&= pb(3d_{\text{h}}+2d) + pd(3d_{\text{h}}+1) \\
M_{\text{embed}} &= M_{\text{input}} + pbsV + W_{\text{embed}} \\
&= pb(d+V) + pdV
\end{align*}

\section{Supplementary Experimental Results}
This section presents comprehensive empirical evaluations to systematically investigate the performance of PIPO.

\subsection{Bandwidth and GPU Utilization}
\label{Appendix:gpu-utilization}
To further investigate the performance improvement provided by PIPO, we measure the transfer speed and GPU utilization, as these factors are crucial in determining inference performance.

\begin{figure}
    \centering
    \begin{minipage}[b]{0.45\textwidth}
        \centering
        \includegraphics[width=\linewidth]{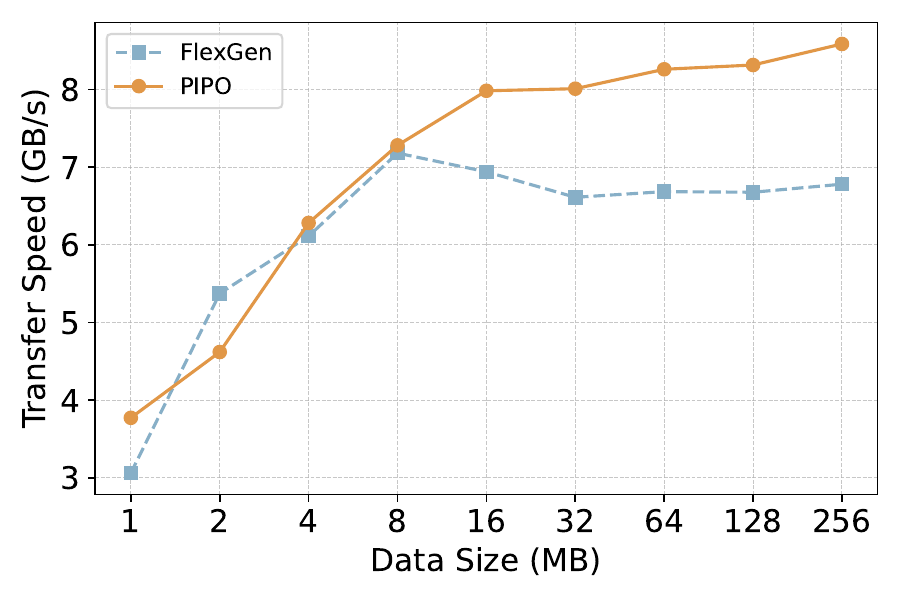}
        \caption{Comparison of data transfer speed.}
        \label{bandwidth}
    \end{minipage}
    \begin{minipage}[b]{0.45\textwidth}
        \centering
        \includegraphics[width=\linewidth]{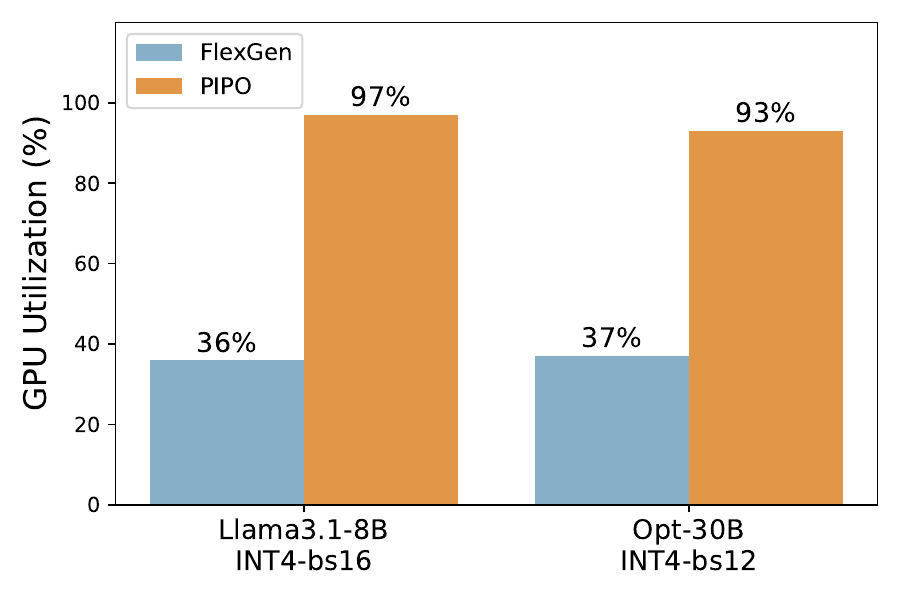}
        \caption{Comparison of GPU utilization.}
        \label{gpu}
    \end{minipage}
\end{figure}

Figure~\ref{bandwidth} shows the transfer speed from disk to GPU memory, corresponding to the weight loading task in PIPO.
Compared to FlexGen which relies on PyTorch's official implementation, PIPO's data transfer suite outperforms it when the data size exceeds 8MB and maintains high efficiency as data size increases. 
Figure~\ref{gpu} presents a comparison of GPU utilization during inference. 
The experiment is conducted on LLaMA3.1-8B model with a batch size of 16 and OPT-30B model with a batch size of 12, both using INT4 quantization and disk-offloading for weights.
Compared with FlexGen, PIPO increases GPU utilization from below 40\% to over 90\%, indicating faster data transfer and more efficient scheduling of the PIPO framework.

\subsection{Ablation Study}
\label{Appendix:ablation}
To analyze the impact of PIPO’s individual optimizations on overall performance, we conduct an ablation study by progressively integrating its key components.

\begin{figure}[h]
  \centering
  \includegraphics[width=0.8\textwidth]{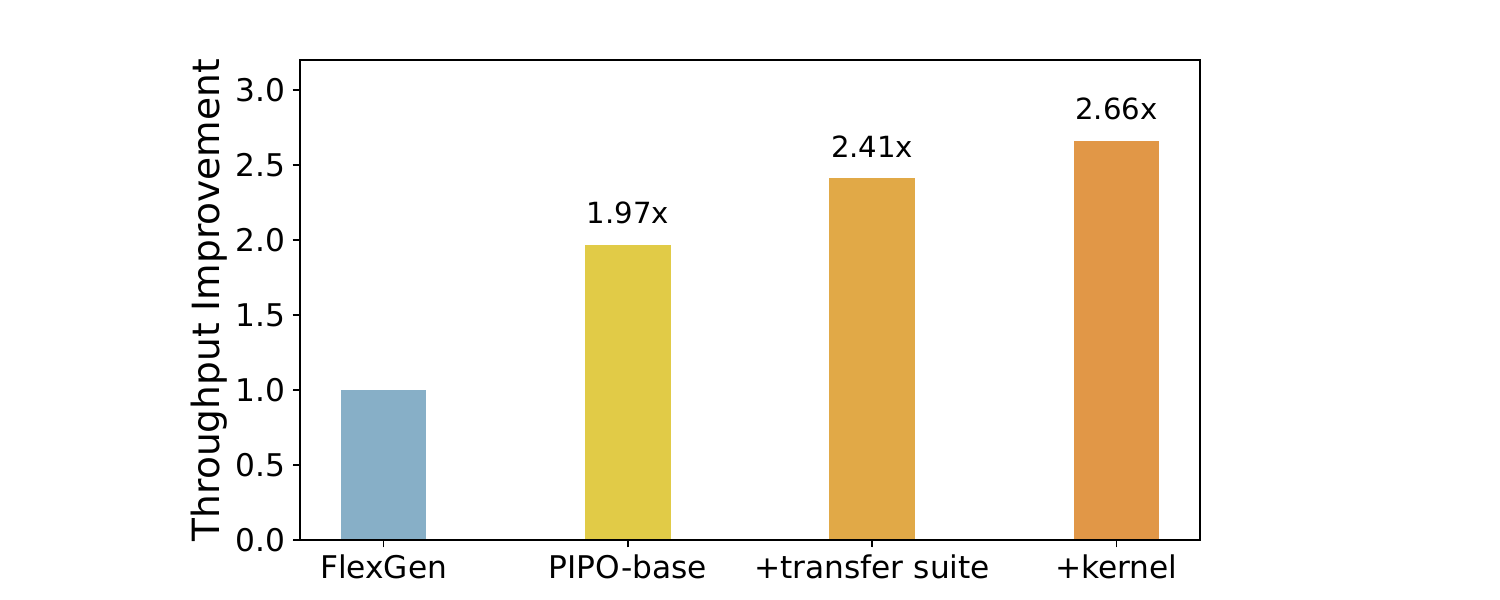}
  \caption{Performance breakdown of PIPO's key components. }
  \label{ablation}
\end{figure}

As illustrated in Figure~\ref{ablation}, PIPO’s pipeline scheduling~(denoted as PIPO-base) alone nearly doubles the throughput of the INT4 OPT-13B model with a batch size of 4 when employing disk-offloading, compared with FlexGen. 
Adding the transfer suite further boosts performance to $2.41\times$, while the compute kernel enhances it to $2.66\times$. 
These results highlight the key factors driving PIPO's efficiency and demonstrate the cumulative benefits of its components. 

\subsection{Extended Context Handling}
\label{Appendix:context-length}
\begin{figure}[h]
  \centering
  \includegraphics[width=0.8\textwidth]{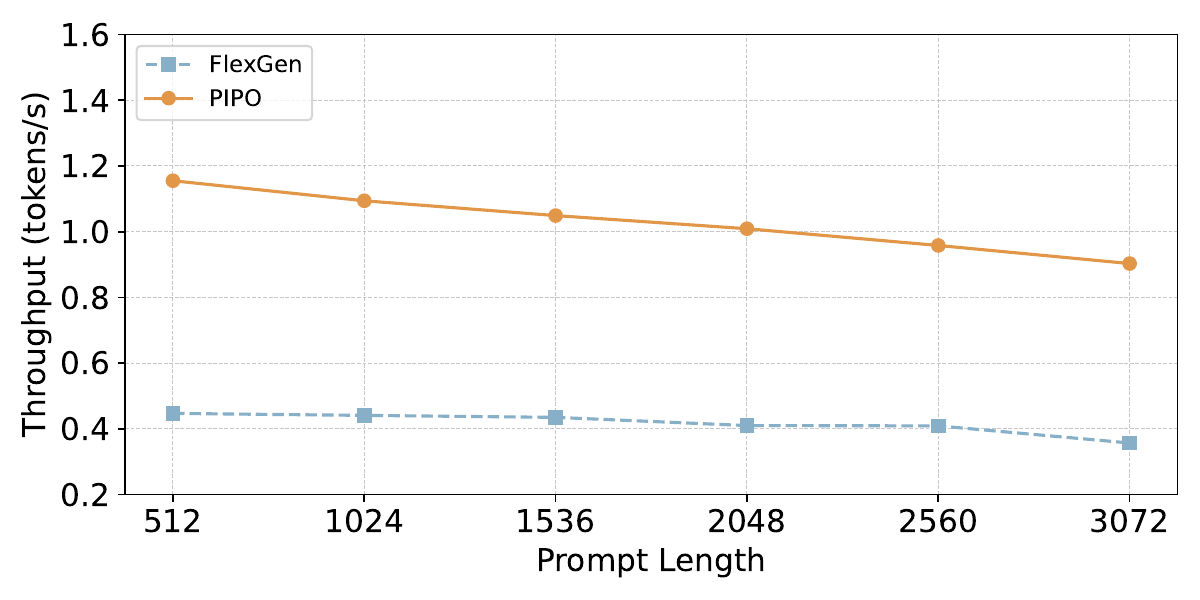}
  \caption{Throughput with variable prompt lengths~(512-3072 tokens). Benchmark conducted on: (1) Hardware: NVIDIA RTX 3060 (6GB VRAM) mobile GPU with SSD disk offloading; (2) Model: LLaMA3.1-8B in INT4 quantized format; (3) Configuration: batch size=1. }
  \label{prompt}
\end{figure}

\begin{figure}[h]
  \centering
  \includegraphics[width=0.8\textwidth]{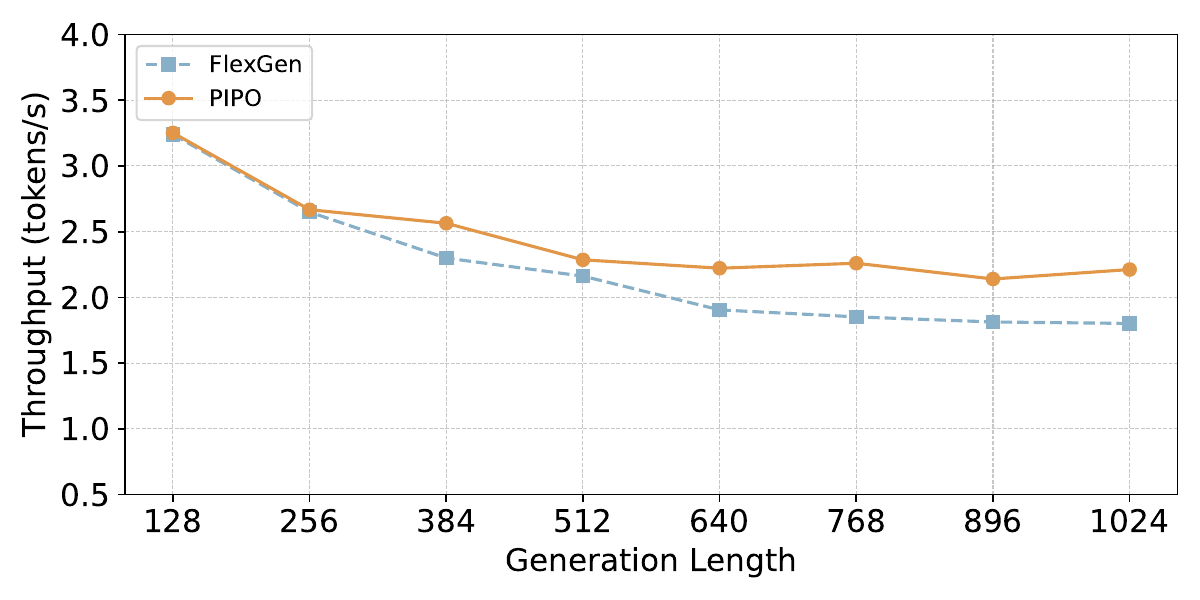}
  \caption{Throughput with variable generation lengths~(128-1024 tokens). Benchmark conducted on: (1) Hardware: NVIDIA RTX 3060 (6GB VRAM) mobile GPU with SSD disk offloading; (2) Model: LLaMA3.1-8B in INT4 quantized format; (3) Configuration: batch size=4. }
  \label{output}
\end{figure}

To evaluate PIPO's capacity in long-form interaction scenarios, we conduct stress tests with progressively expanding prompt and generation length. 
Figure~\ref{prompt} shows PIPO's performance with longer prompt lengths. 
And Figure~\ref{output} shows PIPO's throughput at different generation lengths. 

\subsection{MoE Architecture Benchmark}
\label{Appendix:moe}

Deploying very large models like LLaMA3.1‑405B on consumer devices is challenging due to PCIe bandwidth limits-a 405B model over a 20 GB/s PCIe takes about 40 seconds to traverse every Transformer layer. 
MoE models, on the other hand, offer a promising alternative, as they activate only a subset of experts per token, resulting in more manageable memory usage. 
In MoE models' architecture, usually only a few of the experts will be activated for a input token when decoding.
Mixtral 8x7B has 8 experts, but only 2 experts will be selected for computation.
DeepSeek-R1 has 256 experts but only 8 experts will be selected, which means only 30B parameters are used for a single token's decoding.
This has great advantage for offloading which only loads the needed weight instead of the whole model.
There's still a chanllenge for offloading inference MoE models: we can't predict the experts used before the "gate" operator.
However, there are also opportunities that PIPO can schedule the computation and data transfering:
\begin{itemize}

\item
Cache loading and saving is not affected by this.

\item
In DeepSeek-R1, shared expert is a fixed expert for computation, where we can make some experts' weight loading parallelly here.

\item
When multiple experts are waiting to be loaded and computed, we can overlap one expert's computation with other's weight loading.

\end{itemize}

\begin{figure}[h]
  \centering
  \includegraphics[width=0.8\textwidth]{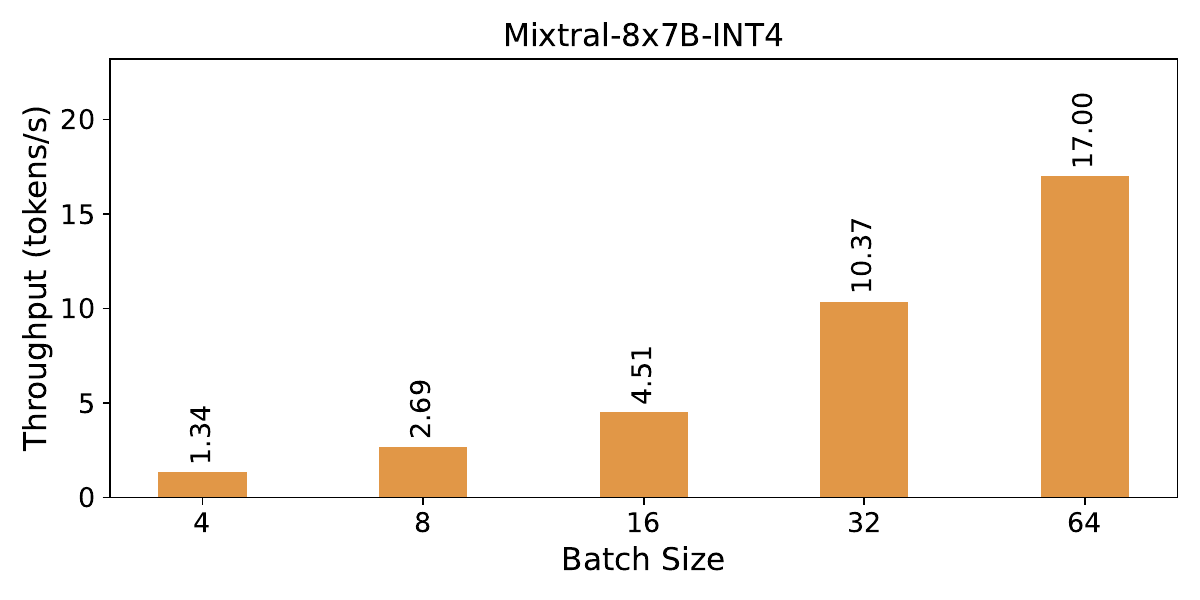}
  \caption{Throughput of Mixtral-8x7B model. Benchmark conducted on: (1) Hardware: NVIDIA RTX 3060 (6GB VRAM) mobile GPU with SSD disk offloading; (2) Configuration: INT4 quantized, cache on VRAM and 512 prompt length. }
  \label{mixtral}
\end{figure}

We have integrated Mixtral 8×7B and DeepSeek-R1 671B into PIPO. 
On an RTX3060 laptop, PIPO achieves a throughput of 12.482 tokens/s on Mixtral 8×7B as shown in Figure~\ref{mixtral} and has managed to finish DeepSeek-R1 671B inference at 0.13 tokens/s, with disk offloading used for both models.
To the best of our knowledge, no existing works have deployed these models on devices with such limited VRAM~(6GB) and DRAM~(16GB).

\subsection{High-end Device Performance}
\label{Appendix:high-end-performance}

We also perform experiments on a desktop with an RTX4090 and 64GB of memory. 
Since smaller models can be loaded directly into GPU memory, we focus on evaluating PIPO using the larger LLaMA3.1-70B and OPT-30B models, as shown in Table~\ref{table_4090_0}. 

\begin{table}[h]
  \caption{Throughput~(tokens/s) on RTX4090}
  \label{table_4090_0}
  \centering
  \begin{tabular}{llll}
    \toprule
    Model              & Weight on   & FlexGen  & PIPO \\
    \midrule
    LLaMA3.1-70B-INT4  & CPU         & 1.113    & 1.214 \\
                       & Disk        & 0.497    & 1.107 \\
    OPT-30B-FP16       & Disk        & 1.113    & 1.214 \\
    \bottomrule
  \end{tabular}
\end{table}

We also store part of the weights on GPU to utilize more of the VRAM and offload the rest weight to CPU, as shown in Table~\ref{table_4090_1}.

\begin{table}[h]
  \caption{Throughput~(tokens/s) on RTX4090 with partial offloading}
  \label{table_4090_1}
  \centering
  \begin{tabular}{llll}
    \toprule
    Model              & Weight on             & FlexGen  & PIPO \\
    \midrule
    LLaMA3.1-70B-INT4  & 40\% GPU, 60\% CPU    & 5.454    & 7.718 \\
    OPT-30B-FP16       & 45\% GPU, 55\% CPU    & 6.870    & 8.665 \\
    \bottomrule
  \end{tabular}
\end{table}

\subsection{Latency Analysis}
\label{Appendix:latency}
Latency performance, particularly time-to-first-token (TTFT) and per-token decoding latency, represents a crucial aspect of inference efficiency. Our evaluation focuses on the LLaMA3.1-8B model (with batch size = 1) under disk offloading conditions, using an NVIDIA RTX 3060 GPU, with results shown in Table~\ref{latency}. 

\begin{table}[h]
  \caption{Latency results on LLaMA3.1-8B (batch size: 1) with disk offloading}
  \label{latency}
  \centering
  \begin{tabular}{lllll}
    \toprule
    & \multicolumn{2}{c}{Time-to-first-token~(s)}  & \multicolumn{2}{c}{Decode latency~(s)}\\
    \cmidrule(r){2-5}
    Context-length     &FlexGen & PIPO   &FlexGen & PIPO \\
    \midrule
    512                & 2.120  & 1.218  & 2.074  & 0.837 \\
    1024               & 2.371  & 1.717  & 2.027  & 0.842 \\
    1536               & 2.686  & 2.336  & 2.031  & 0.857 \\
    2048               & 3.406  & 2.956  & 2.266  & 0.876 \\
    \bottomrule
  \end{tabular}
\end{table}

\subsection{Offloading Overhead Analysis}
\label{Appendix:overhead}
While offloading is widely adopted to address high memory demands in LLM inference, this technique may negatively impact computational efficiency. 
For models that can be accommodated within consumer-grade device memory, PIPO strategically disables offloading to maximize throughput.

To systematically quantify offloading impacts, we conduct experiments applying both disk and CPU offloading strategies to memory-sufficient models. 
This comparative analysis reveals the offloading overhead, demonstrating how PIPO's optimized disk offloading mechanism narrows the performance gap with CPU offloading.
Key results are presented in Tables~\ref{llama32_1b}, \ref{llama31_8b} and~\ref{memory}.

\begin{table}[h]
  \caption{Throughput Benchmark Comparison (LLaMA3.2-1B)}
  \label{llama32_1b}
  \centering
  \begin{tabular}{llrr}
    \toprule
    & & \multicolumn{2}{c}{Throughput~(tokens/s)} \\
    \cmidrule(r){3-4}
    Weight Location    & Cache Location & FlexGen & PIPO \\
    \midrule
    GPU                & GPU            & 132.770 & 130.716 \\
                       & CPU            & 127.392 & 128.551 \\
    CPU                & GPU            & 32.666  & 32.636 \\
                       & CPU            & 31.227  & 31.233 \\
    Disk               & GPU            & 3.852   & 16.982 \\
                       & CPU            & 3.622   & 16.307 \\
    \bottomrule
  \end{tabular}
\end{table}

\begin{table}[h]
  \caption{Throughput Benchmark Comparison (LLaMA3.1-8B)}
  \label{llama31_8b}
  \centering
  \begin{tabular}{llrr}
    \toprule
    & & \multicolumn{2}{c}{Throughput~(tokens/s)} \\
    \cmidrule(r){3-4}
    Weight Location    & Cache Location & FlexGen & PIPO \\
    \midrule
    CPU                & GPU            & 5.497  & 6.058 \\
                       & CPU            & 5.449  & 6.009 \\
    Disk               & GPU            & 1.926  & 4.925 \\
                       & CPU            & 1.765  & 4.695 \\
    \bottomrule
  \end{tabular}
\end{table}

\subsection{Memory Footprint Profiling}
\label{Appendix:memory}

PIPO achieves memory efficiency similar to FlexGen, albeit with some key differences. Table~\ref{memory} detailed memory usage results for OPT-6.7B (INT4) with a prompt length of 256 and a batch size of 4.

\begin{table}[h]
  \caption{Memory Usage and Throughput Benchmark Comparison between FlexGen and PIPO}
  \label{memory}
  \centering
  \begin{tabular}{lllrrr}
    \toprule
    Weight Location & Cache Location & Framework & \multicolumn{1}{r}{Throughput} & \multicolumn{1}{r}{VRAM Usage} & \multicolumn{1}{r}{DRAM Usage} \\
    & & & \multicolumn{1}{r}{(token/s)} & \multicolumn{1}{r}{(GB)} & \multicolumn{1}{r}{(GB)} \\
    \midrule
    GPU & GPU & FlexGen & 7.184 & 5.424 & 8.90 \\
     &  & PIPO & 9.174 & 5.666 & 6.30 \\
    \cmidrule{2-6}
     & CPU & FlexGen & 6.837 & 5.166 & 7.90 \\
     &  & PIPO & 9.172 & 5.342 & 7.30 \\
    \midrule
    CPU & GPU & FlexGen & 7.089 & 1.796 & 16.39 \\
     &  & PIPO & 8.151 & 1.904 & 15.18 \\
    \cmidrule{2-6}
     & CPU & FlexGen & 6.460 & 1.438 & 16.57 \\
     &  & PIPO & 8.028 & 1.486 & 16.49 \\
    \midrule
    Disk & GPU & FlexGen & 2.276 & 1.896 & 9.28 \\
     &  & PIPO & 6.729 & 2.012 & 6.35 \\
    \cmidrule{2-6}
     & CPU & FlexGen & 2.122 & 1.438 & 9.75 \\
     &  & PIPO & 6.428 & 1.522 & 7.28 \\
    \bottomrule
  \end{tabular}
\end{table}

Under same settings, PIPO consumes 200 MB more VRAM due to preloading, but its transfer suite reduces DRAM usage by 2GB. 
With disk offloading, PIPO achieves the same throughput as FlexGen (which offloads weights to CPU DRAM) while cutting DRAM usage by up to 10 GB. 
Compared to a non-offload implementation, PIPO reduces VRAM usage by 66.4\% with an 11.2\% performance degradation.

\section{Combination with Parallelism Techniques}
\label{Appendix:parallelism}
PIPO can be extented to multi-GPU by combining with parallelism techniques. 
Data Parallelism (DP), Tensor Parallelism (TP), and Pipeline Parallelism (PP) can be seamlessly integrated into PIPO by simply modifying the weight loading process, specifically the loaded content and its destination.
Assuming one new layer is going to be loaded during offloading inference: 
\begin{itemize}
    \item DP forces every GPU to load the layer
    \item TP lets each GPU load its portion of the layer
    \item PP requires only one GPU to load the layer
\end{itemize}

With PCIe bandwidth as the bottleneck, inference speed hinges on the data loading latency. 
As a result, when extending PIPO to multi-GPU setup, DP performs the worst, while TP achieves the best performance when GPUs have isolated PCIe bandwidth. 
High-end consumer motherboards typically offer isolated PCIe channels for GPUs, whereas lower-end PCs may share bandwidth. Servers usually provide isolated PCIe channels.

Additionally, Expert Parallelism (EP), as seen in DeepSeek-R1, can also be integrated into PIPO. 
It distributes experts within a MoE layer across GPUs and forwards tokens to their corresponding experts during inference. 
Essentially, it's similar to applying TP to an entire MoE layer but only loads the required experts instead of the whole layer, making it friendly to pipeline offloading.

\end{document}